\documentclass{osa-article}
\usepackage{braket}
\newcommand{\fla}[1]{\begin{flalign}#1\end{flalign}}
\journal{oe}

\articletype{Research Article}

\begin{document}

\title{Plug\&Play subcarrier wave quantum key distribution with deep modulation}

\author{Oleg I. Bannik\authormark{1} 
and E. S. Moiseev\authormark{1,2,*}}

\address{\authormark{1}Kazan Quantum Center, Kazan National Research Technical University, 18a Chetaeva str.,  \protect\\ Kazan, 420111, Russia\\
\authormark{2}KQT LLC,  41-022 Galiaskar Kamal str., Kazan, 420021, Russia \\
\authormark{*}e.s.moiseev@kazanqc.org} 



\begin{abstract}
We report a quantum key distribution using subcarrier wave encoding in Plug\&Play configuration. 
Our detailed study shows how subcarrier encoding operates in a presence of Rayleigh back-scattering, chromatic dispersion and other experimental negative factors that occur in Plug\&Play configuration.
We propose and experimentally demonstrate a novel regime for the subcarrier wave encoding, where simultaneous detection of two logical states in a single basis is combined with robustness and simplicity of the original protocol. This deep modulation regime shows reliable quantum communication with losses up to 20 dB and promises a twice increase in secret key generation rate in comparison with the original subcarrier wave quantum key distribution in one-way or Plug\&Play configurations.
\end{abstract}

%
\section{Introduction}
Nowadays, quantum key distribution (QKD) is one of the most developed quantum technologies. 
In QKD, two authenticated parties, commonly referred as Alice and Bob, create secure keys by exploiting quantum properties of light. Modern conventional cryptography is based on the assumption of computational complexity and is vulnerable to an attacker with sufficient computing power, while QKD combined with a one-time pad encryption technique is able to provide a way for achieving unconditional  confidentiality of information transmission
 \cite{Gisin2002}. 
Over the last two decades a huge effort has been carried out to build commercial QKD systems \cite{8745822,article}.

One of the promising practical QKD encoding methods is subcarrier wave (SCW) phase encoding \cite{PhysRevLett.82.1656}.
SCW QKD originally is a one-way system and it is notable for its interferometer-free optical scheme and robustness to environmental effects on a fiber line \cite{PhysRevLett.82.1656,Gleim:16}. Meanwhile, Plug\&Play configuration \cite{Ribordy2000, PPMDI, Valivarthi:20} is convenient for several applications such as client-server QKD networks \cite{BOGDANSKI2009258}, practical SCW twin-field implementations \cite{Chistiakov:19} and creating a cheaper client device. 
The system in the Plug\&Play configuration uses the double-pass scheme: Bob sends a reference laser beam to Alice, which she subsequently modulates to encode information and redirects it back to Bob. Double-pass scheme is a way to compensate for polarization distortions in an optical fiber and to avoid complex active compensation. Plug\&Play configuration allows to combine most of the bulky and expensive parts like a laser source, an optical filter, detectors and optical switches on the Bob's side. Alice's client device can be greatly simplified, resulting in reduced deployment cost and improved scalability of a such quantum client-server network.
Thus evaluation of SCW QKD in Plug\&Play configuration is of certain practical interest.

Rayleigh backscattering from the strong reference beam is one of the key problems in QKD systems with the double-pass optical design.  It  considerably increases false counts and hence limits communication distance \cite{Subacius2005}. Several methods have been proposed to circumvent this problem such as an introduction of time separation between  signal and Rayleigh backscattered radiation through the storage line \cite{Stucki_2002}, use of a separate fiber line to deliver the reference radiation \cite{Valivarthi:20}, or use of several spectral channels and sequential switching between them \cite{Peng_2008}.
These methods successfully help to reduce negative effect of backscattered radiation, but at the cost of limiting the communication speed, or increasing complexity and overall cost of a QKD setup. Considered SCW type of encoding is less susceptible to negative effects of Rayleigh backscattering. This is due to the fact that signal sideband photons carrying encoded information are partially frequency separated from the back-scattered photons and there is also no need to send powerful optical pulses except for the carrier providing the required power level at the output of Alice's device.

Here we theoretically analyze and experimentally implement SCW QKD in Plug\&Play configuration. 
First we develop a mathematical model of SCW quantum key distribution in the Plug\&Play configuration.
We derive an expression for secret key generation for the given set of control parameters taking into account different practical aspects such as losses, Rayleigh backscattering, chromatic dispersion, and non-ideal optical filters and detectors. 
Next, we propose and analyze a new regime of SCW QKD, namely deep modulation, which allows to increase key generation rate twice. 
The key improvement is an ability to distinguish two logical states in a single measurement in the deep-modulation regime in contrast to one state in the original scheme \cite{Gleim:16}.
Then we find the optimal control parameters for Plug\&Play SCW QKD depending on the communication distance. 
Finally, we describe the experimental setup and perform an experimental SCW QKD with deep modulation. 
At the end, we discuss the results and perspective of Plug\&Play SCW QKD and the proposed new regime. 



\section{Mathematical description}
\subsection{Basic model}

In the Plug\&Play configuration, Bob provides an optical reference signal to Alice via an optical channel with power attenuation coefficient $\eta$. 
It consists from a train of coherent-state pulses with duration $T$ each, carrier frequency $\omega_0$ and complex amplitude $\alpha_0$.  
Alice encodes information by applying a sinusoidal phase-modulation with modulation depth $\beta$, modulation frequency $\Omega$, and phase $\phi_A$ to the received reference light. 
An optical state after Alice's phase modulator is described by wavefunction
\fla{
\ket{\psi} = \bigotimes^{\infty}_{m=-\infty} \ket{ \sqrt{\eta} \alpha_0 J_{m}(\beta)e^{i m \phi_A} }_m, \label{eq::Bob-State}
}
where multiplication is carried over $m\in \mathbb{Z}$ indices of sidebands with frequencies $ \omega_0 + i m\Omega$, $J_{m}(\beta)$ is m-th order Bessel function of the first kind.

Following BB84-family of protocols \cite{BB84}, a single basis set is formed by pairs of states \eqref{eq::Bob-State} with $\phi_A$ being shifted by $\pi$ with respect to each other. 
To implement BB84 protocol at each clock cycle Alice chooses random value for basis and logical bit. 
It is done by randomly picking $\phi_A$ either from  $\{``0"\rightarrow 0,``1"\rightarrow \pi\}$ or from $\{``0"\rightarrow\pi/2,``1"\rightarrow 3\pi/2\}$ sets. Here a right arrow indicates correspondence between logical state marked by braces and phase to be chosen to the right of the arrow.

Alice sends state  \eqref{eq::Bob-State} to Bob by reflecting it back through the same lossy fiber. 
To decode Alice's bit at each clock cycle, Bob applies phase modulation to his optical channel with the same modulation depth $\beta$, modulation frequency $\Omega$ and his own phase $\phi_B$.
After that, Bob separates signal on the sidebands from a carrier frequency by sending signal through an optical frequency filter with extinction ratio $\rho$. 
The subsequent photon detection in the sidebands reveals the encoded information, if Bob has sifted both basis and phases $\phi_A=\phi_B$ correctly.

Probability of the detector to click is
\fla{
P_{\text{det}}(\phi_a,\phi_b) = \left( \eta_D \frac{n_{\text{ph}}(\phi_A,\phi_B) }{T} +\gamma_{\text{det}} +   \eta_D \frac{n_{\text{rs}}}{T} \right)\delta t, \label{eq:pdet}
}
where $\eta_D$ is a quantum efficiency of a single photon detector,  $\delta t$ is its gating time and $\gamma_{\text{det}}$ is a dark count rate, $n_{\text{ph}}$ and $n_{\text{rs}}$ are average numbers of Alice's  and Rayleigh back-scattered photons, that reach Bob's detector, respectively.
To quantify the number of photons that reaches Bob's detector $n_{\text{ph}}$, 
we consider two situations where Bob's phase equals to Alice's one ($\phi_A=\phi_B$) or differs by $\pi$ ($\phi_A=\phi_B+\pi$), whereas we assume that Bob has sifted the basis correctly.
Ideally, if Bob's and Alice's phases coincide, the sidebands are populated as if it was a double modulation, otherwise all photons are demodulated into the carrier.

A finite extinction ratio of the frequency filter leads photons on a carrier frequency being detected. 
We model filter as a Bragg grating mirror, that reflects only a carrier frequency ($m=0$) with power coefficient $r$ and transmits with power coefficient $\tau$. Meanwhile, photons on the sidebands ($m\neq0$) are transmitted through the filter with power coefficient $1-\varrho$ and reflected with power coefficient $\varrho$, where $\varrho$ is called as filter's sideband suppression factor.
Taking into account this effect, the average number of Alice's photons at Bob's detector is
\fla{
n_{\text{ph}} (\phi_A,\phi_B) = |\alpha_0|^2 \eta^2 \cdot \begin{cases}
 (1-\varrho)(1-J^{2}_{0}(2\beta))+ \tau J^{2}_{0}(2\beta) \quad &\phi_A = \phi_B \\
 \tau \quad &\phi_A = \phi_B+\pi \\
\end{cases}
\label{eq:nph},
}
where we assume negligibly small jitter between Alice's and Bob's modulation oscillators over a repetition period. 

The detector catches not only signal photons arriving from Alice, but as well noisy photons generated by Rayleigh backscattering $n_{\text{rs}}$. 
Their amount is mainly determined by an intensity of the reference beam $\alpha_0$, 
length of the fiber and Rayleigh backscattering coefficient ($\beta_{\text{RS}} \approx -40$ dB for a single-mode fiber) \cite{Subacius2005}. Portion of these photons is frequency shifted into modulation sidebands by Bob's phase modulation, that reduces the number of detected noise photons by factor of $1- J^{2}_{0}(\beta)$.
The resultant number of photons from Rayleigh backscattering at Bob's detector is
\fla{
n_\text{rs} \approx \left( (1-\rho) (1- J^{2}_{0}(\beta)) + \tau J^{2}_{0}(\beta)  \right)(1-\eta^2 )
\beta_{\text{RS}} |\alpha_0|^2. \label{eq:nrs}
}

Next, we simulate secret key generation rate against beam splitter collective attack. 
Secret key rate $K$ is lower limited by Devetak-Winter bound \cite{Devetak2005,Miroshnichenko:18}: 
\fla{
K = \nu P_B (1-  f(Q) H(Q) - \chi(A:E) ), \label{eq:keyrate}
}
where $\nu=1/T$ is a repetition rate and $P_B$ is probability of bit detection at Bob's side per pulse.  
Probability $P_B$ can be represented as a total probability of detecting a photon with correct and wrong bit decoding. Namely it is a probability that Alice encodes logical bit value x and Bob get the same bit value x or opposite bit value: $(x+1)\text{mod}~2$. For $x=0$ it is $P_B=(P_{\text{det}}(0,0)+P_{\text{det}}(0,\pi))/2$, where factor $1/2$ comes from the 2 bases used in BB84 protocol. 
The product $f(Q)H(Q)$ defines an amount of information revealed during error correction with $f(Q)$ being an efficiency of error correcting protocol and $H(Q)$ being a binary Shannon entropy:
\fla{
H(Q) = -Q \log_2 (Q) - (1-Q) \log_2(1-Q).
}
Here $Q$ is a quantum bit error probability (QBER), which is a ratio of detection with false decoding to the probability of detection $P_B$:
\fla{
Q = \frac{P_{\text{det}}(0,\pi)}{P_{\text{det}}(0,0)+P_{\text{det}}(0,\pi)}. \label{eq:qber}
}

The term $\chi(A:E)$ is an upper bound for an amount of information that is accessible to an eavesdropper Eve for a given collective attack. 
It was shown \cite{Miroshnichenko:18} that a collective beam splitter attack provides the following information to Eve:
\fla{
\chi(A:E) = H\left(\frac{1}{2}\left(1-e^{-|\alpha_0|^{2}\eta (1-\eta) (1-J_0(2\beta)) } \right)\right).\label{eq:chi}
}

In practice, there are different losses in a QKD system that do not contribute to Rayleigh backscattering \eqref{eq:nrs} such as insertion losses of the optical elements, monitoring photodiodes, etc.  
We include these losses into the model by dividing overall losses $\eta =\eta_A \eta_B \eta_F$ into single pass losses in the Alice's encoder $\eta_A$, losses in Bob's detection path $\eta_B$ and losses in optical fiber $\eta_F$. Next we assume, that Rayleigh backscattering is associated only with fiber-related losses $\eta_F$. It is done by replacing $(1-\eta^2)$  in Eq.~\eqref{eq:nrs} with $(1-\eta_F^2)\eta_B$. 
The single pass in Bob's detection path and double pass in fiber and Alice's device are attributed by replacing $\eta^2$ with $\eta^{2}_A \eta^{2}_F \eta_B$ in Eq.~\eqref{eq:nph}.
We assume Alice's and Bob's devices are trusted, i.e. photons that are lost in the devices are not intercepted by Eve.
In this case, the mutual information in Eq.~(7) is modified by replacing $\eta(1-\eta)$ with  $\eta^{2}_A\eta_F(1-\eta_F)$.
\
Eventually, we estimate the key rate by first combining Eq.~\eqref{eq:nph} and Eq.~\eqref{eq:nrs} in  Eq.~\eqref{eq:pdet}, that is used to calculate QBER \eqref{eq:qber}. 
The resulted QBER entropy and  eavesdropper's mutual information from Eq.~\eqref{eq:chi} is used to calculate the key rate bound in Eq.~\eqref{eq:keyrate}. 

\subsection{Chromatic dispersion}

So far we have ignored an effect of chromatic dispersion on the QKD performance. 
It is a valid approximation if dispersion compensator is used on Alice's side.  Otherwise the dispersion should be included into analysis at distances longer than 25 km \cite{Kiselev:20}.

The chromatic dispersion in optical fibers manifests itself as nonlinear frequency-dependant phase shift. 
It is usually expressed 
as the Taylor expansion of a wavenumber in the vicinity of central frequency $\omega_0$ \cite{boyd2020nonlinear}:
\fla{
k(\omega-\omega_0) =  k_0 +  \frac{\partial k}{\partial \omega} \bigg\vert_{\omega_0} \left(\omega-\omega_0\right)  + \frac{1}{2}\frac{\partial^2 k}{\partial^2 \omega}\vert_{\omega_0} \left(\omega-\omega_0\right)^2 + ... 
}
The values $\frac{\partial k}{\partial \omega}=1/v_g$ and $\frac{\partial^2 k}{\partial^2 \omega}=\text{GVD}$ are usually expressed as inverse group delay ($1/v_g$) and group velocity dispersion ($\text{GVD}$) respectively. 
Further on we assume value of GVD to be $-2.0407 \cdot 10^{-23}$ $\text{s}^{2}/$km as in G.652 single mode fiber.

In SCW QKD, where Alice sends a wavepacket with equidistantly separated sidebands, the chromatic dispersion modifies the state \eqref{eq::Bob-State} sent by Alice to Bob by an additional nonlinear phase modulation acquired over distance $d$:
\fla{
\ket{\Psi} \rightarrow \ket{\Psi'(\beta,\phi_A)} = 
\bigotimes^{\infty}_{m=-\infty} \ket{ \eta_A\eta_F \sqrt{\eta_B} \alpha_0 J_{m}(\beta)e^{i m \phi_A + i m \Omega (1/v_g + m\cdot\text{GVD}\cdot\Omega/2)  d } }_m,
\label{eq:stateCH}
}
where distance $d$ is related to the fiber losses as $\eta_F = d \cdot 0.2 \text{dB/km}$.
The effect of inverse group delay is easily compensated by an adjusting phase difference between Alice's and Bob's modulation signals by $d\Omega/v_g $. 
The contribution of quadratic effect is not that straightforward, since opposite sideband pairs accumulate destructive phase difference at different distance \cite{Kiselev:20}.
We include an effect of GVD on the secret key rate generation by replacing simplified value of average number of photons in Eq.~\eqref{eq:nph} with:
\fla{
n'_{\text{ph}}(\phi_A,\phi_B) = (1-\rho)\bra{\Psi'(\beta,\phi_A)}U^{\dagger}(\beta,\phi_B) \left( \sum_{m\neq 0} a^{\dagger}_m a_m \right) U(\beta,\phi_B)\ket{\Psi'(\beta,\phi_A)}   \nonumber \\
+ \tau \bra{\Psi'(\beta,\phi_A)}U^{\dagger}(\beta,\phi_B) a^{\dagger}_0 a_0  U(\beta,\phi_B)\ket{\Psi'(\beta,\phi_A)},
\label{eq:new-photon}
}
where $a^{\dagger}_m$ ($a_m$) is a photon creation (annihilation) operator at $m$'th sideband, $U(\beta,\phi_B)$ is beam-splitter-like unitary operator representing an action of Bob's phase modulator such that
\fla{
U^{\dagger}(\beta,\phi_B) a_k  U(\beta,\phi_B) = \sum^{\infty}_{l=-\infty} J_{l}(\beta) a_{k+l} e^{il\phi_B}.
\label{Eq::ModulatorUnitary}
}
It is worth noting, that in general more sophisticated model of electro-optical modulator \cite{Capmany:10,Miroshnichenko:17} should be used as infinite number of indices $l$ leads to nonphysical sidebands with negative frequencies.  However for practical modulators and moderate values of modulation depths ($\beta<4$) simple model as in  Eq.~(\ref{Eq::ModulatorUnitary}) is a good approximation as we can limit summation to the first five pairs of sidebands which contain more than 99\% of total energy in all sidebands.
By substituting Eq.~\eqref{eq:new-photon}  instead of  Eq.~\eqref{eq:nph} in Eq.~\eqref{eq:pdet},  we calculate the key rate bound by Eq.~\eqref{eq:keyrate} similarly as in case without chromatic dispersion.

\subsection{Deep modulation}
In original SCW QKD systems, a strong laser beam is phase modulated by Alice to produce weak sidebands with given phases, that encode quantum information with respect to 0-th order of modulation. 
The measurement is performed as single photodetection at sidebands wavelengths only \cite{PhysRevLett.82.1656}. 
In the case of using deep modulation regime where $\beta=\beta_{\text{DM}} \approx 1.2$ such that $J_0(2\beta_{\text{DM}}) = 0$, detection can be carried out on both carrier and subcarrier wavelengths. 
Thus, it is easy to see, that the proposed regime allows to distinguish two phases $\phi_A=\phi_B$ and $\phi_A=\phi_B+\pi$ during a single measurement. 

Let Bob choose phase $\phi_B$ at each clock cycle out of two values  0 or $\pi/2$. 
If Bob picked the Alice's basis, all optical energy is transferred into the sidebands when $\phi_A=\phi_B$. 
However, if $\phi_A=\phi_B+\pi$ the energy is completely transferred into the carrier.  
For small amplitudes $|\alpha_0|^2 J^{2}_0(\beta_{\text{DM}}) < 1$ presence of a photon at carrier or sideband frequencies can be detected by placing single photon counting modules at carrier and sidebands channels after the spectral filter, that separates carrier from the sidebands. 
Therefore, Bob can distinguish $\phi_A$ in given basis without an explicit match as in the original SCW encoding. 
At a proper intensity level, the protocol is capable of performing QKD with roughly twice faster key generation rate as we show below.

We still use Eq.~\eqref{eq:keyrate} to bound key generation rate. 
The only things we recalculate are QBER and $P_b$ Eq.~\eqref{eq:qber}, as now we can distinguish two logical states. 
We find these values by first introducing two identical detectors Dc and Dsb, that detect photons after reflection from the filter (carrier mode) or after transmission trough the filter (sideband's modes) respectively. 
The probability of getting a click for the detector Dsb is similar to Eq. \eqref{eq:pdet}:
\fla{
& P_{\text{Dsb}}(\phi_A,\phi_B)  =
 \bigg(\frac{\eta_D \tau}{T} \bra{\Psi'(\beta_\text{DM},\phi_A)}U^{\dagger}(\beta_{\text{DM}},\phi_B) a^{\dagger}_0 a_0  U(\beta_{\text{DM}},\phi_B)\ket{\Psi'(\beta_{\text{DM}},\phi_A)}  \nonumber \\
 & + \frac{\eta_D (1-\rho)}{T}  
\bra{\Psi'(\beta_\text{DM},\phi_A)}  
U^{\dagger}(\beta_{\text{DM}},\phi_B) \bigg( \sum_{m\neq 0} a^{\dagger}_m a_m \bigg) U(\beta_{\text{DM}},\phi_B)
\ket{\Psi'(\beta_{\text{DM}},\phi_A)} 
 \nonumber \\
 & + \gamma_{\text{Det}} + \frac{\eta_D}{T}n_{\text{rs}}
\bigg)\delta t, 
}
in turn, probability of detecting photon at the carrier frequency is 
\fla{
&P_{\text{Dc}}(\phi_A,\phi_B) = \bigg(  \frac{\eta_D r}{T} \bra{\Psi'(\beta_\text{DM},\phi_A)}U^{\dagger}(\beta_{\text{DM}},\phi_B) a^{\dagger}_0 a_0  U(\beta,\phi_B)\ket{\Psi'(\beta_{\text{DM}},\phi_A)}   
\nonumber\\
&+ \frac{\eta_D\varrho}{T}  \bra{\Psi'(\beta_{\text{DM}},\phi_A)}U^{\dagger}(\beta_{\text{DM}},\phi_B) \left( \sum_{m\neq 0} a^{\dagger}_m a_m \right) U(\beta_{\text{DM}},\phi_B)\ket{\Psi'(\beta_{\text{DM}},\phi_A)}  + \frac{\eta_D}{T}  n_{\text{rs}}' +\gamma_{\text{Det}}\bigg) \delta t.
}
Here the number of noise photons reaching detector Dc due to Rayleigh backscattering $n_\text{rs}'$ differs from Eq.~\eqref{eq:nrs} by replacing
$(1-\rho) (1- J^{2}_{0}(\beta))$ and $\tau J^{2}_{0}(\beta))$ with $r J^{2}_0(\beta_{\text{DM}})$ and $\rho (1-J^{2}_0(\beta_{\text{DM}}))$, respectively.

Similarly as in Eq.~\eqref{eq:qber}, we calculate QBER as a ratio between probability of erroneous bit
decoding to total probability of bit decoding from Alice to Bob
\fla{
Q = \frac{ P_E } {P_C + P_E },
\label{eq:dm-qber}
}
with probability of erroneous bit decoding is  
\fla{
P_E=P_{\text{Dsb}}(\phi_A, \phi_A+\pi )(1-P_{\text{Dc}}(\phi_A, \phi_A+\pi)) + P_{\text{Dc}}(\phi_A, \phi_A )(1-P_{\text{Dsb}}(\phi_A, \phi_A ))
}
and probability of correct bit decoding is:
\fla{
P_C=P_{\text{Dsb}}(\phi_A, \phi_A )(1-P_{\text{Dc}}(\phi_A, \phi_A )) + P_{\text{Dc}}(\phi_A, \phi_A+\pi )(1-P_{\text{Dsb}}(\phi_A, \phi_A+\pi )).
}
The derived QBER \eqref{eq:dm-qber} together with Eq.~\eqref{eq:chi} is used in  Eq.~\eqref{eq:keyrate} to estimate key generation rate in deep modulation regime.
For the beam splitter attack in deep modulation regime the amount of information accessible to Eve is still quantified by Eq.\eqref{eq:chi} as we show in the appendix.

The deep modulation regime can be straightforwardly used with decoy states method to counter photon number splitting attack \cite{Hwang03,Lo05, Wang05}, if Alice applies amplitude modulation and optical phase randomization to the pulses at her site.
The asymptotic secret key rate can be estimated by modified Gottesman, Lo, Lutkenhaus, and Preskill bound \cite{Lo05, Ma05}:
\fla{
K = \frac{\nu}{2}\left(  P_1 \left( 1- H(Q_1) \right) - P_{\alpha} f(Q) H(Q)    \right) ,
}
where $P_{\alpha}$ is a fraction of detection events per pulse (gain) due to signal coherent state with given amplitude $\alpha$,
$P_1$ is a fraction of single photon detection events, 
$Q_1$ is a QBER for single photon component. 
The later two values are estimated using decoy states method as described in \cite{Ma05}.
However in our experiment we were not able to implement decoy states method due to electronic hardware's constrains. Thus for a sake of consistency in differential comparison of original SCW and deep modulation regime  further on we consider key generation rate in both regimes in a presence of the collective beam-splitter attack.

\subsection{Optimization}

\begin{table}[htp]
\centering
\begin{tabular}{|c|c|}
\hline
Parameter & Description \\ \hline
$\eta_D=0.1$ & Detector quantum efficiency  \\ \hline
$\Omega=2\pi \cdot$ 5 GHz & Modulation frequency \\ \hline
$\gamma_{\text{det}}=50$ Hz &  Detector's dark count rate \\ \hline
$\tau=0.01$ &  Filter's carrier transmittance \\  \hline 
$r=0.99$ & Filter's carrier reflection \\ \hline
$\varrho=-38$ dB & Filter's sideband suppression factor \\ \hline 
$\text{GVD}=-2.0407 \cdot 10^{-23}$ $\text{s}^{2}/$km 
& Group velocity dispersion  \\ \hline 
$1/v_g = 4.9\cdot10^{-6}$  s/km & Inverse group delay  \\ \hline 
$\delta t=$3.3 ns & Gating time \\ \hline 
$T=$10 ns & Pulse repetition period \\ \hline 
$\eta_A=-6$dB & Alice's device losses\\ \hline 
$\eta_B=-6$dB & Bob's device losses \\ \hline 
\end{tabular}
\caption[Parameters used to find the noise spectrum]{Model parameters }
\label{table:parameters}
\end{table}

\begin{figure}[h!!]
\includegraphics[width=6.2cm]{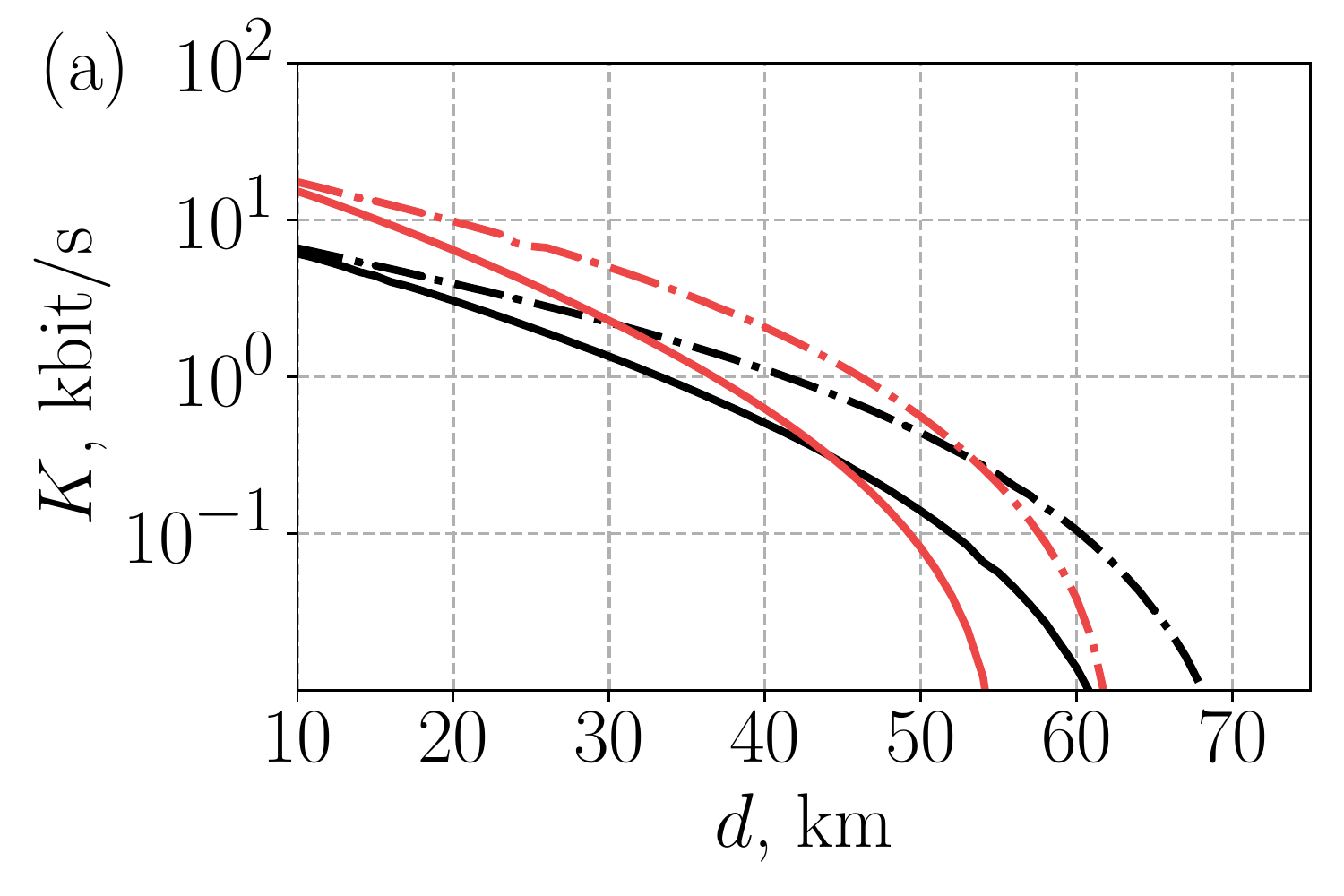} \quad
\includegraphics[width=6.2cm]{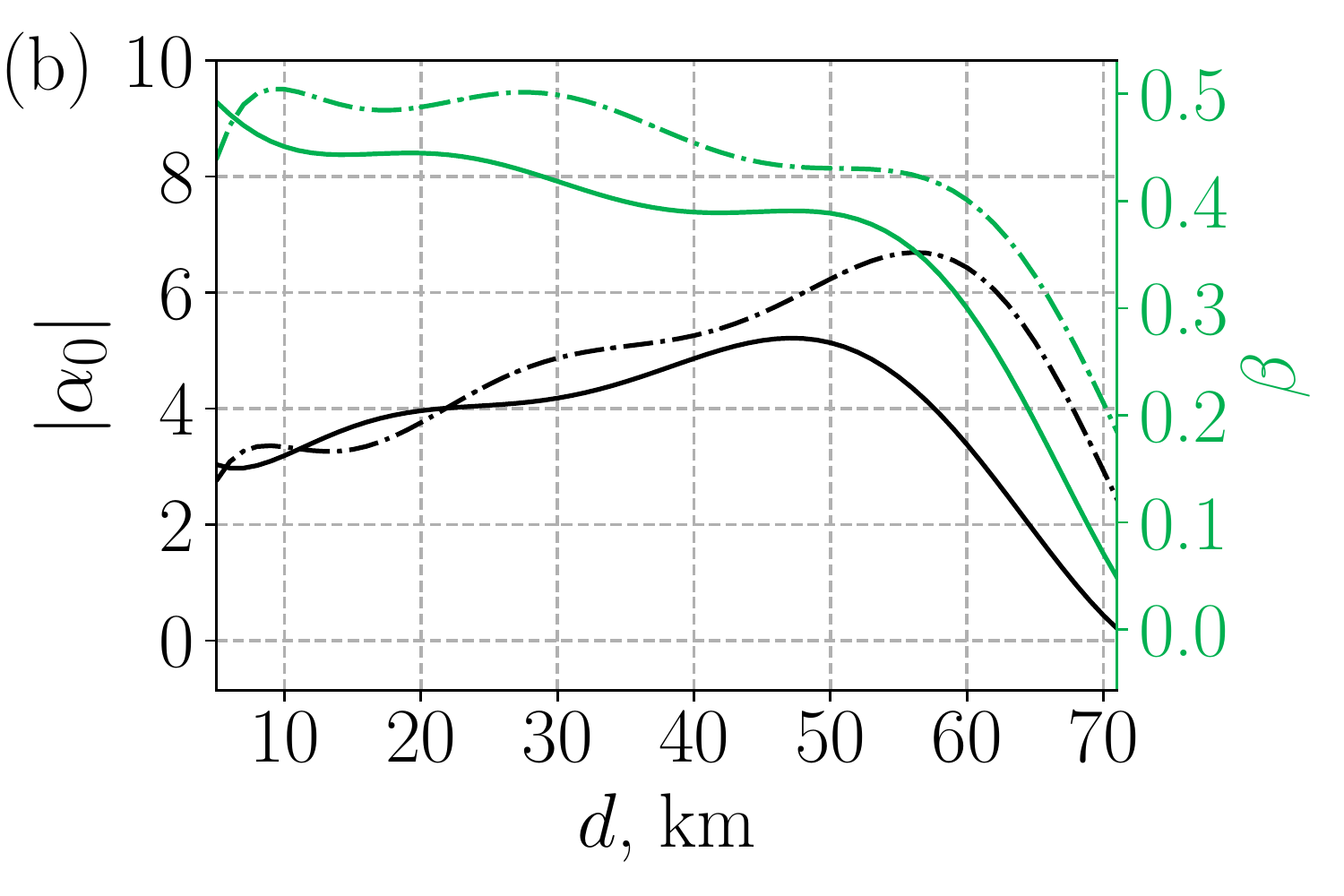}
\caption{(a) Key generation rate $K$ vs communication distance $d$  at optimal parameters with losses in Alice's and Bob's devices $\eta_A=\eta_B=-6$dB
for original protocol with (\textbf{black} dot-dashed) and without (\textbf{black} solid) dispersion compensation. $K$ in deep modulation regime with (\textcolor{red}{red} dot-dashed) and without (\textcolor{red}{red} solid) dispersion compensation.
(b) The optimal values of $|\alpha_0|$ (\textbf{black}) and $\beta$ (\textcolor{green}{green}) vs single pass power loss $\eta_F$. Dot-dashed  and solid lines differentiate cases with and without dispersion compensations respectively.
}
\label{Fig::theory-device-loss}
\end{figure}

We numerically optimize the key generation rate \eqref{eq:keyrate} at given $\eta_F$  for the
original and proposed deep modulation regime of SCW QKD in Plug\&Play configuration. 
In our calculation we use experimentally accessible parameters, which are summarized in Table \ref{table:parameters}.  
We consider Alice and Bob both having identical losses $\eta_A=\eta_B=-6$ dB in their devices and assume moderate efficiency of reconciliation protocol $f(Q)=1.25$.
In Fig.~\ref{Fig::theory-device-loss} (a) we present a key generation rate for optimal control parameters with and without  dispersion compensation. 

As it was expected the deep modulation regime increases key generation rate twice at low and moderate losses with and without dispersion compensation. 
Similarly to one-way SCW QKD communication \cite{Kiselev:20}, the chromatic dispersion decreases communication distance in Plug\&Play configuration. 
Especially chromatic dispersion effects the deep modulation regime, where more energy is distributed in high-order sidebands ($m>1$), which experience stronger dispersion. 
Eventually at approximately 60 km dispersion completely suppresses key generation rate, while original protocol could be capable of delivering few hundreds of secret bit/s.  

In Fig.~\ref{Fig::theory-device-loss} (b) we present the optimal control parameters for the original protocol with and without dispersion. 
Without dispersion compensation, optimal $\beta$ tends to be around $0.38 \pm 0.05$ as it is depicted in Fig.~\ref{Fig::theory-device-loss}(b). 
This value of modulation depth is a compromise between reduction of the Rayleigh backscattering and increase of QBER due to chromatic dispersion. 
Use of dispersion compensation removes an increase of QBER with larger $\beta$.  
In turn, it allows to increase $\alpha_0$ at higher losses without dramatic increase of QBER by Rayleigh backscattering. 
At deep modulation,  $\alpha_0$ is a single control parameter, since $\beta$ is fixed. 
We find that the optimal $\alpha_0$ does not significantly change over a full range of $\eta$ and lies within $1.5\pm 0.1$  region.


\begin{figure}[h!]
\centering\includegraphics[width=6cm]{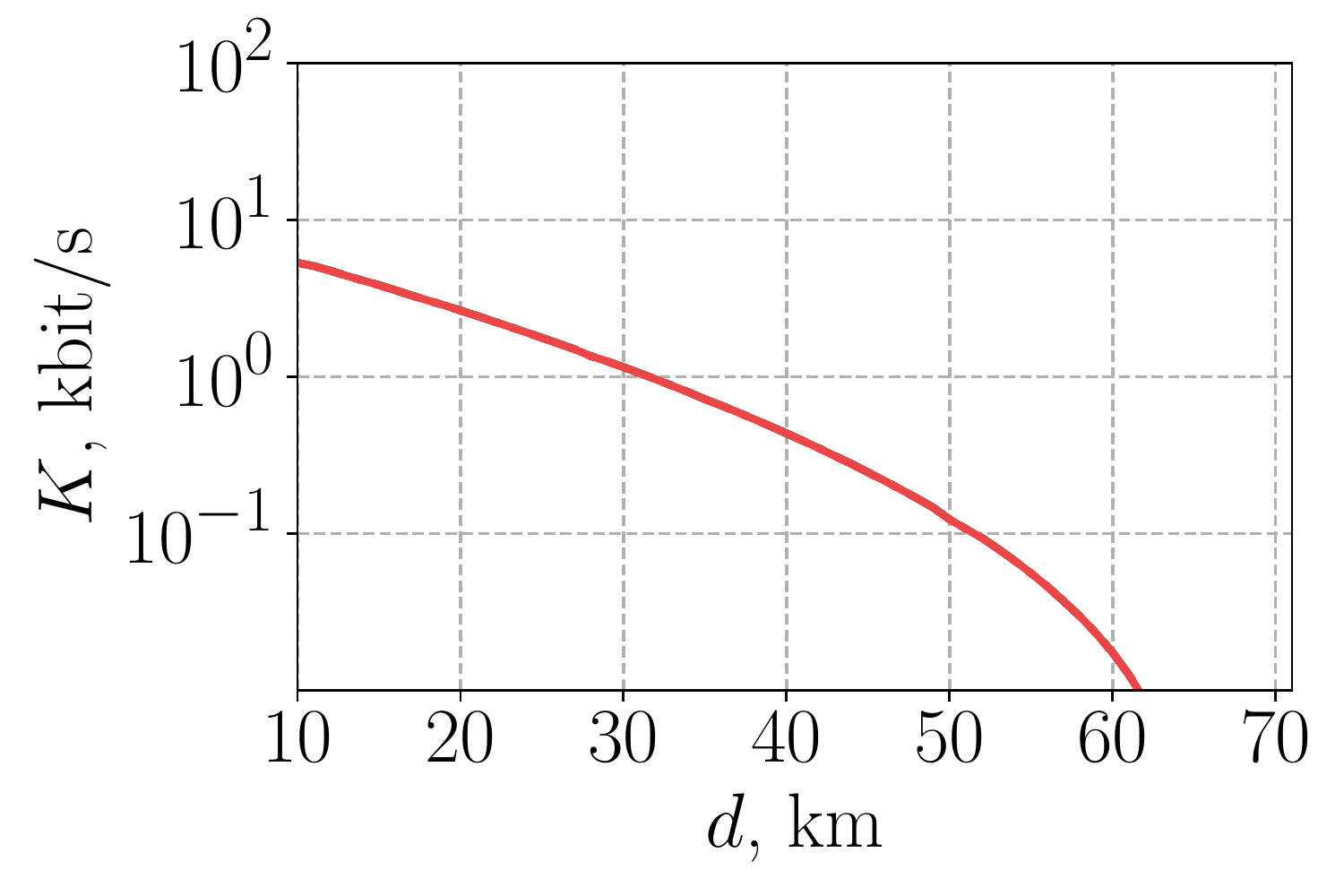}
\caption{ $K$ for deep modulation regime with $\eta_A=-5$ dB and $\eta_B=-7.5$ dB and carrier detection or original protocol with $\eta_A=-5$ dB and $\eta_B=-10.5$ dB. The device losses are chosen for our experimental setup. 
}
\label{Fig::comparison}
\end{figure}

Despite that developed model determines SCW in deep modulation with dispersion compensation as an optimal protocol at moderate distances, there are numbers of practical issues interfering with this conclusion.
The typical solutions for compensating dispersion at Alice's side are Bragg gratings or dispersion compensating fiber. 
The latter option is cumbersome  with significant losses. 
The Bragg gratings could be designed with little extra losses, but with major overhead in the final price of the setup. 
The extra detector would also increase final price in comparison with single detector configuration.
In turn, detecting single photon at the sidebands may require additional optical filter to protect the detector from stray light.
Extra filter would introduce additional losses ( $\sim -3$ dB for typical in-fiber design) into Bob's device, that would decrease key rate proportionally.  
There is no such problem for SCW QKD in deep modulation regime with only a single detector monitoring carrier, where spectral filter \textit{in situ} removes stray light.
We compare the key generation rate of SCW QKD in deep modulation with single detector and original protocol with extra device loss of 3 dB, that simulates additional filter.  
We find no difference between these two protocols in terms of key rate generation as present in Fig.~\ref{Fig::comparison}.  
Next we experimentally implement Plug\&Play SCW QKD in deep modulation with single detector.

\section{Experiment}
\begin{figure}[h!]
\centering\includegraphics[width=12cm]{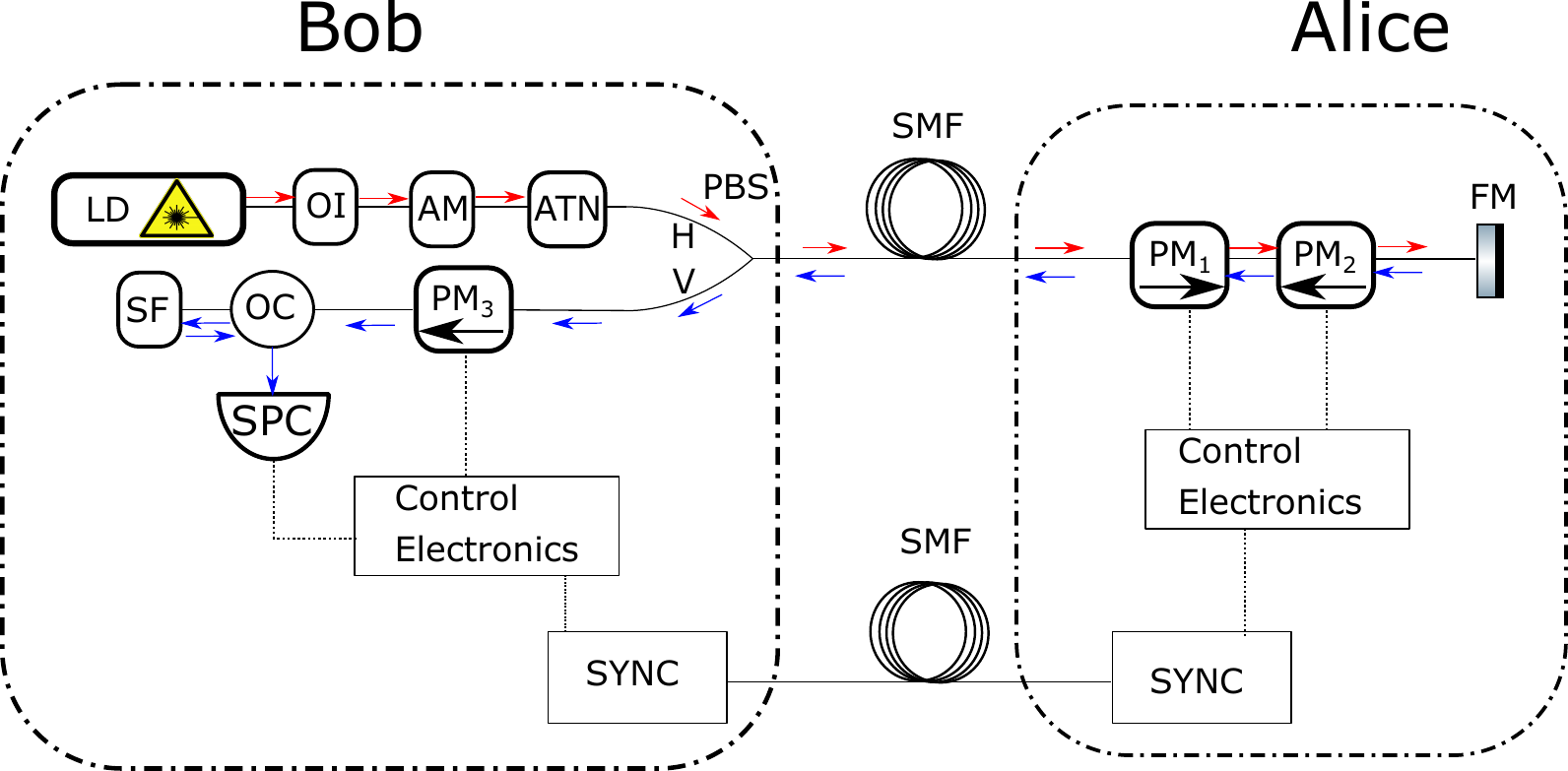}
\caption{ Principle scheme of SCW Plug\&Play QKD system, that consists from laser diode (LD), optical isolator (OI), amplitude modulator (AM), optical attenuator (ATN), polarization beam splitter (PBS), single mode fiber line (SMF), 
phase modulators (PM), Faraday mirror (FM), optical circulator (OC), spectral filter (SF),  modulation phase synchronization unit (SYNC), single photon counting module (SPC). Here \textcolor{red}{red} arrow indicates the direction of reference light emitted by Bob, \textcolor{blue}{blue} arrows indicate the direction of Alice's photons with encoded information, \textbf{black} arrow within PM indicates a direction of optical signal  for which phase modulation is possible.
}
\label{Fig::scheme}
\end{figure}

The proposed scheme uses the same basic  components of one-way SCW QKD system \cite{Gleim:16}, whereas we convert it into Plug\&Play configuration as  depicted in Fig.~\ref{Fig::scheme}. 
Here the laser source is a continuous-wave fiber Bragg grating stabilized laser diode with linewidth < 1 MHz operating at a wavelength tunable within C-band. 
The wavelength is adjusted to match the spectral filter reflection band.
The laser diode is protected by an optical isolator, while variable optical attenuator is used to set carrier power to the necessary level taking into account the losses in the fiber-optic line.  
$\text{LiNb}\text{O}_3$ Mach-Zehnder amplitude modulator is used to generate 3.3 ns pulses with a repetition rate of 100 MHz to concentrate the signal photons inside the detection gate clocked at 100 MHz.
Prepared reference radiation is routed through the H-port of a three-port fiber polarization beam splitter (PBS) to its common port, which is connected to Alice via 25-km single mode fiber link with optical losses of 5 dB.

Alice uses special composite phase-modulation unit to make communication polarization insensitive. 
It consists out of two polarization sensitive $\text{LiNb}\text{O}_3$ phase modulators  $\text{PM}_1$ and $\text{PM}_2$ with the same principal axis but opposite direction of microwave tracks. 
During the first pass only a single polarization component is phase modulated by $\text{PM}_1$, where direction of microwave and optical signals coincides. 
After reflection from the Faraday-mirror, the polarization is flipped and orthogonal polarization component is modulated by $\text{PM}_2$. 

Each PM is driven by 5 GHz sine-wave signal generator. 
The relative phase between Alice's and Bob's drivers is dynamically adjusted by synchronization circuit (SYNC). 
SYNC consists of the phaselock loop unit that locks the phase of 5 GHz generator to 100 MHz logical clock. 
Small form-factor pluggable optical modules are used to distribute the clock from Bob to Alice via separate fiber link. 
Similar phase lock unit at Alice's side adjusts the phase of her 5 GHz signal generator.  
At each 100 MHz clock cycle Alice's and Bob's control electronics randomly choose phases states of their 5 GHz microwave signals from a set of four shifted by $\pi/2$ to encode information and measurement basis respectively.
Alice's and Bob's modulation bases are synchronised automatically within each key generation cycle. 

Since the signal is returned to Bob over the same fiber path, polarization distortion is compensated and thus, linearly polarized signal is routed to the V-port of PBS. 
Next, after the signal package is phase modulated again at $\text{PM}_3$, the carrier is separated from sidebands on 5 GHz Fiber Bragg Grating spectral filter and directed to the single photon counter (SPC) by an optical circulator. 
In the experiment, we applied registration only at carrier wavelength due to the limitations of our current hardware, which supports only one SPC. 
Photon counting is performed by an avalanche photodiode-based SPC (ID Quantique ID210) with detection efficiency $\eta_D=0.1$,  detection window $\delta t \approx 3.3$ ns, dead time $10$ $\mu$s resulting in dark count rate of $\gamma_{\text{det}}=50$ Hz.
The data was collected in blocks of about 2500 bits and errors were corrected by LDPC-codes \cite{Elkouss09}.

\begin{figure}[h!]
\includegraphics[width=6.2cm]{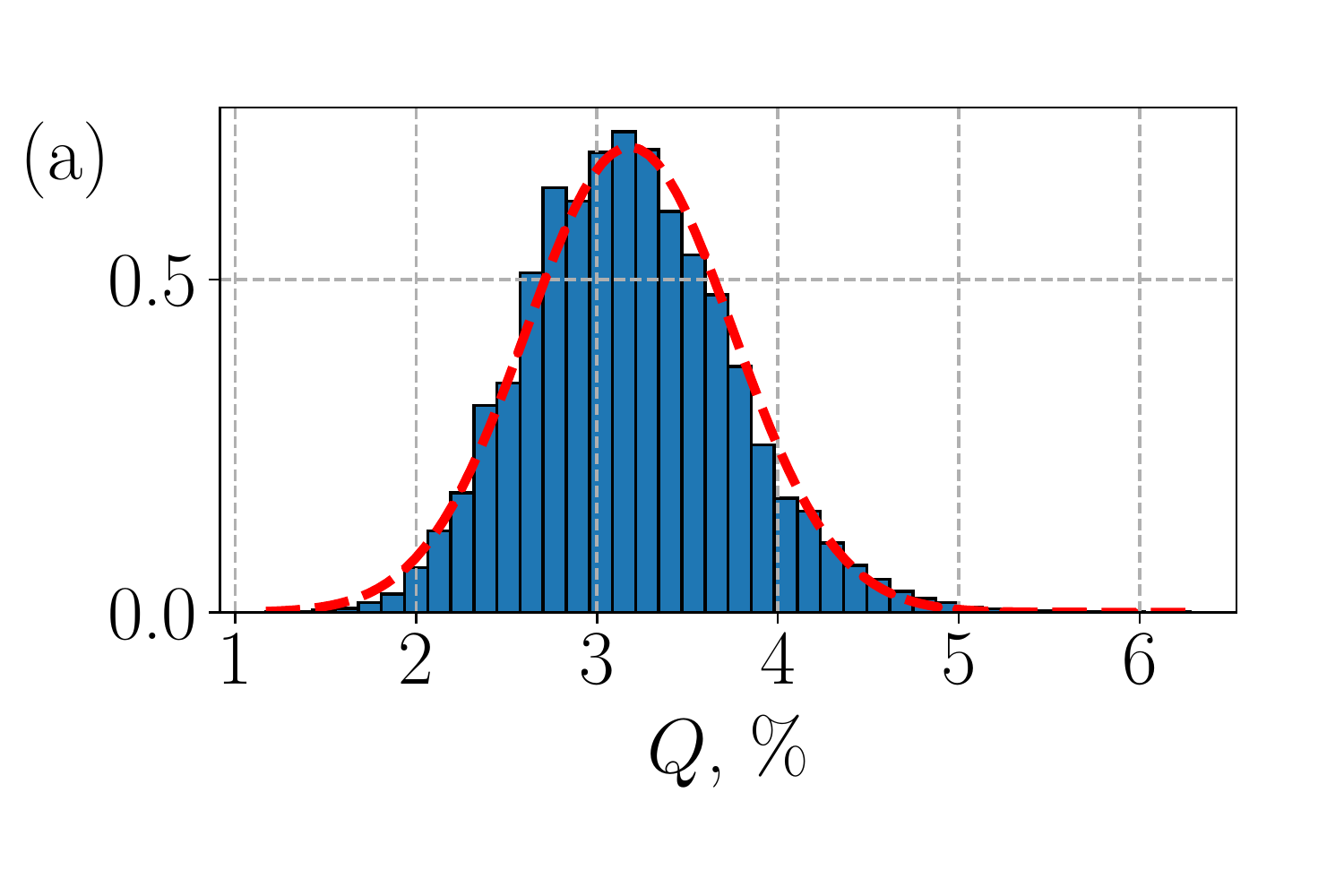} \quad
\includegraphics[width=6.2cm]{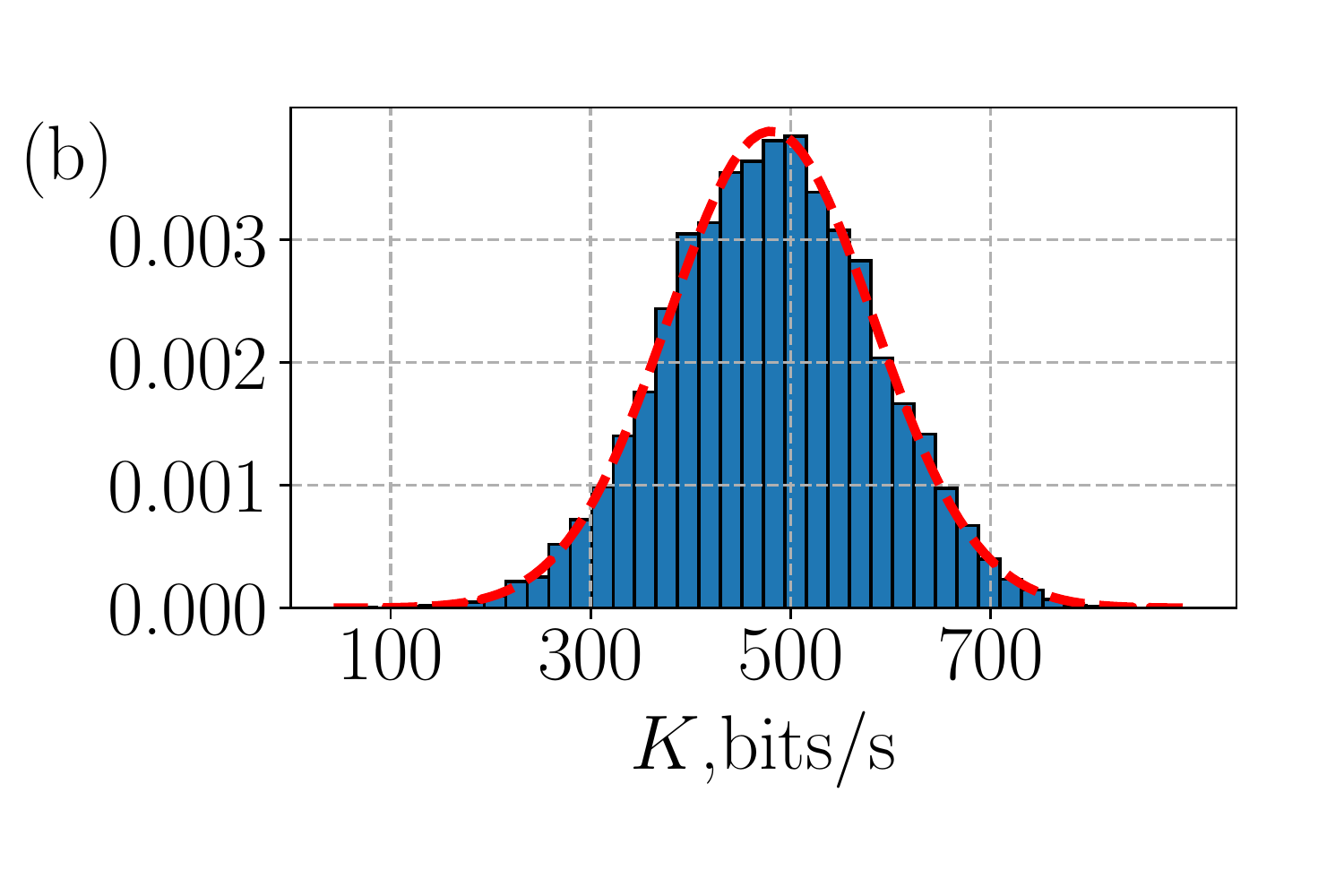}
\caption{ Normalized histograms of QBER (a) and secret key rate (b) acquired during 24 hours of continuous communications.}
\label{Fig::exp-data}
\end{figure} 
 
Single pass losses from Alice's Faraday mirror to Bob's SPC is 17.5 dB. They consists from 5 dB losses at Alice's phase modulators (2.5 dB each), 5 dB and 7.5 dB losses in 25 km fiber spool and in Bob's detection path respectively. 
The latter losses are composed from 3 dB, 2.5 dB and 2 dB insertions losses of optical circulator, $\text{PM}_{3}$ and spectral filter respectively.

We performed a 24-hour quantum key distribution in Plug\&Play configuration with deep modulation and single photon detection on the carrier wavelength. 
Alice and Bob were connected by 25 km optical fiber with loss of 5 dB. 
No chromatic dispersion compensation were employed.
We achieved a mean secret key rate of approximately 480 bits/s with average QBER of 3.18\%. 
Both values show Gaussian statistics with respective root mean square of  102 bits/s and 0.32\% as presented in Fig.~\ref{Fig::exp-data}(a) and Fig.~\ref{Fig::exp-data}(b).
The performed test showed robustness of the communications as temperature in the test area was varying by 10 $^{\text{o}}$C over 24 hours.

\section{Conclusion and Discussion}
In summary, we proposed and performed first experimental   subcarrier wave quantum key distribution in Plug\&Play configuration. 
We conducted thoughtful  analysis of the main factors affecting on QKD performance like non-ideal devices, Rayleigh backscattering and chromatic dispersion.  The obtained experimental results indicate that SCW QKD in Plug\&Play configuration with deep modulation can be successfully applied at moderate distances for intracity networks.
Moreover the proposed deep modulation regime can increase communication speed 2 times in SCW QKD.  

It worth noting, that  few  points should be addressed in any future implementations of the proposed SCW QKD. 
It is well known that Plug\&Play QKD systems are particularly susceptible to Trojan horse attacks on Alice's modulator \cite{Gisin2006}. To address this issue the signal, before getting to Faraday mirror, can be sampled by 50-50 non-polarizing beam splitter and measured by 2-channel single photon detection module (omitted in experimental scheme for implementation simplicity). The count and coincidence rates have to be monitored to ensure single photon level and exclude Trojan Horse attack.
For deep modulation regime an absence of the strong reference \cite{Koashi04,Tamaki09, Gleim:16} would require a  use of decoy state method against photon-number-splitting attack \cite{Lo05,Wang05}. It can be achieved by an extra amplitude modulation of the reference beam at Alice's site. 
The minimization of the losses in Alice's and Bob's devices is an important future task, that could be solved in on-chip fashion \cite{Sibson2017, Bunandar18}.

\section*{Appendix: Holevo bound for deep modulation regime}
Here we consider collective beam splitter attack in deep modulation regime.
As in analysis for original protocol \cite{Miroshnichenko:18} we assume, that Eve intercepts all photons lost in the optical channel between Alice and Bob and records them into her quantum memory unit. For sake of simplicity we assume that Eve ``emulates'' losses by inserting beam splitter with power reflectively coefficient $1-\eta_F$ at the beginning of the optical channel. Eve records the reflected photons into her memory, while she sends the transmitted photon to Bob via lossless optical channel.
For a given value of Alice's phase $\phi$ the wavefunction of Eve memory unit is 
\fla{
\ket{\psi_E(\phi)}  = \bigotimes^{\infty}_{m=-\infty} \ket{ \eta_A\sqrt{(1-\eta_F)}\alpha_0 J_m(\beta_{\text{DM}}) e^{im\phi} }.\label{eq::QM}
}
However $\phi$ is yet unknown to Eve, thus before the measurement Eve stores even mixture of the state \eqref{eq::QM} with phases $\phi\in\{0,\pi/2,\pi,3\pi/2\}$ in her memory unit. Once Alice and Bob finish basis sifting and announce the basis Eve has to distinguish two $\pi$-phase shifted states in a fully mixed state composed out of these two states. Next we consider states with phases 0 and $\pi$ as similar analysis can be applied to set of $\pi/2$ and $3\pi/2$, which are converted to 0 and $\pi$ by a proper unitary rotation \cite{Miroshnichenko:18}:
\fla{
\rho = \frac{1}{2} \ket{\psi_E(0)} \bra{\psi_E(0)} + \frac{1}{2} \ket{\psi_E(\pi)} \bra{\psi_E(\pi)}.
}
The information that is accessible to Eve from density matrix $\rho$ being a mixture of two pure states is value of von Neumann entropy of $\rho$ \cite{Hol73}. 
In turn, the von Neumann entropy can be calculated as a Shannon entropy of eigenvalues of operator $\rho$. The eigenvalues of $\rho$ are
\fla{
\lambda_{1,2} = \frac{1}{2}\left( 1 \pm |I(0,\pi)| \right),
}
where $I(\phi_1,\phi_2)$ is an overlap between two states:
\fla{
I(\phi_1,\phi_2) = \bra{ \psi_E (0) }\psi_E (\pi) \rangle.
}
It is calculated using an expression for scalar product of two coherent states:
\fla{
\langle \alpha | \beta \rangle  = \exp \left( -\frac{1}{2}\left( |\alpha|^2 + |\beta|^2\right) + \alpha^* \beta \right),
}
that leads to the following expression for the overlap 
\fla{
I(\phi_1,\phi_2) = \exp \left( -|\alpha_0|^2\eta^{2}_A (1-\eta_F) \sum^{\infty}_{m=-\infty} \left( |J_m(\beta_{\text{DM}})|^2 (1-e^{im(\phi_1-\phi_2)}) \right) \right),
}
for $\phi_1=0,\phi_2=\pi$ we obtain 
\fla{
I(0,\pi) = \exp \left( -|\alpha_0|^2\eta^{2}_A (1-\eta_F)\left(1-J_0(2\beta_{\text{DM}})\right)  \right) = \exp \left( -|\alpha_0|^2\eta^{2}_A (1-\eta_F) \right),
}
where we used the properties of Bessel function of first kind:
\fla{
\sum^{\infty}_{m=0} J^{2}_{2m+1}(\beta_{\text{DM}}) =  \frac{1-J_0(2\beta_{\text{DM}})}{2} = \frac{1}{2}.
}
Thus the Holevo bound is
\fla{
\chi(A:E) = H\left(\frac{1}{2} \left( 1 - e^{-|\alpha_0|^2\eta^{2}_A (1-\eta_F)} \right)  \right).
}

\begin{backmatter}

\bmsection{Acknowledgments}
Authors thank S.A. Moiseev and A. Tashchilina for fruitful discussions.
We are grateful to K. S. Melnik for technical assistance. 
Authors appreciate support within framework project \# 00075-02-2020-051/1 from 02.03.2020.
\bmsection{Disclosure} The authors declare no conflicts of interest. 
\bmsection{Data availability}
Data underlying the results presented in this paper may be obtained from the authors upon reasonable request.
\end{backmatter}
\bibliography{SCW}

\end{document}